\documentclass[journal]{IEEEtran}

\usepackage{cite}
\usepackage{amsmath,amssymb,amsfonts}
\usepackage{algorithmic}
\usepackage{graphicx}
\usepackage{textcomp}
\usepackage[table]{xcolor}
\usepackage{booktabs}
\usepackage{multirow}
\usepackage{hyperref}
\usepackage{tikz}
\usepackage{pgfplots}
\pgfplotsset{compat=1.18}
\usepackage{subcaption}

\begin{document}
\title{\raisebox{-0.3ex}{\includegraphics[height=1.2em]{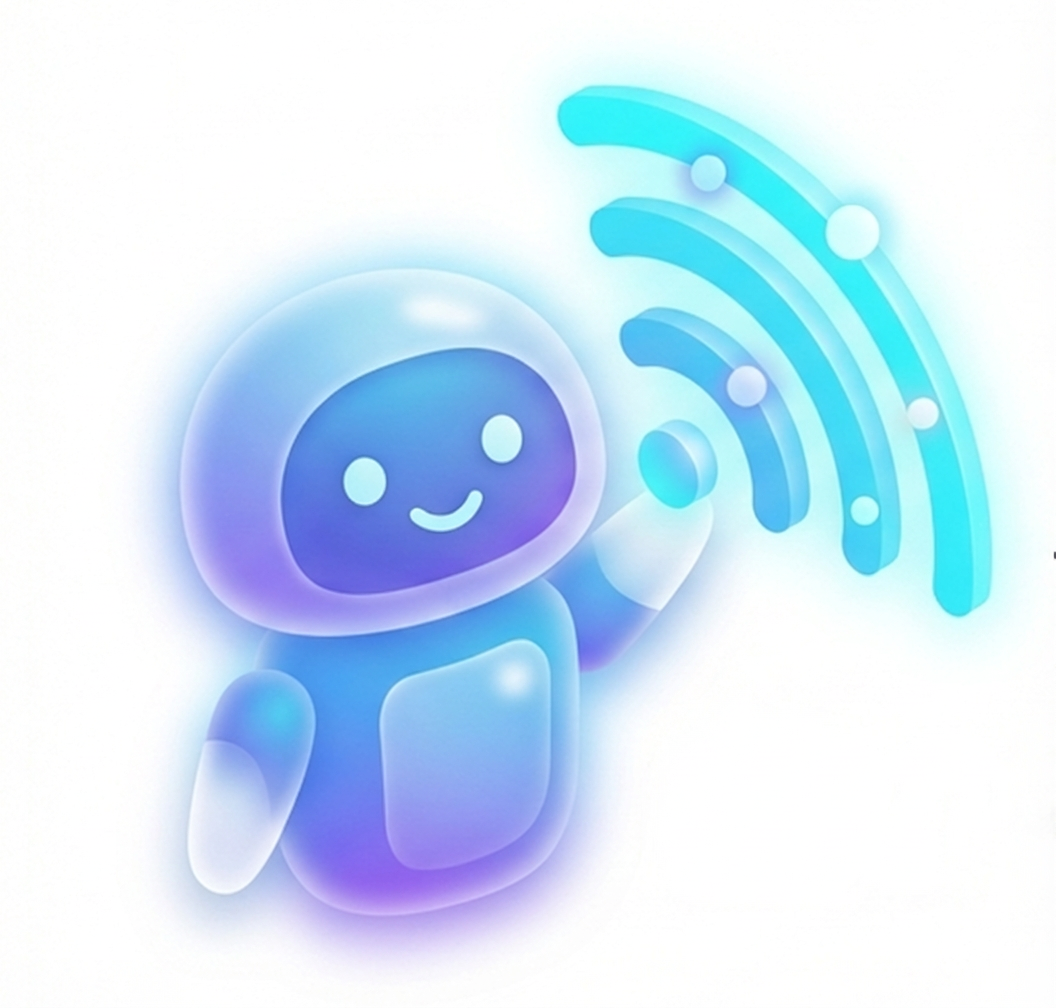}}\,6GAgentGym: Tool Use, Data Synthesis, and Agentic Learning for Network Management}

\author{
  \IEEEauthorblockN{Jiao Chen\IEEEauthorrefmark{1},
  Jianhua Tang\IEEEauthorrefmark{1},
  Xiaotong Yang\IEEEauthorrefmark{2},
  and Zuohong Lv\IEEEauthorrefmark{3}}\\
\IEEEauthorblockA{\IEEEauthorrefmark{1}Shenzhen Smart City Technology Development Group Company, Ltd.}\\
\IEEEauthorblockA{\IEEEauthorrefmark{2}Shien-Ming Wu School of Intelligent Engineering, South China University of Technology}\\
\IEEEauthorblockA{\IEEEauthorrefmark{3}China Unicom Group Co., Ltd.}
\thanks{Corresponding author: Jianhua Tang (jtang4@e.ntu.edu.sg). J. Chen (202110190459@mail.scut.edu.cn) and J. Tang are with Shenzhen Smart City Technology Development Group Company, Ltd. X. Yang (202510192907@mail.scut.edu.cn) is with the Shien-Ming Wu School of Intelligent Engineering, South China University of Technology.
Z. Lv (lvzh67@chinaunicom.cn) is with China Unicom Group Co., Ltd.}
}

\maketitle

\begin{abstract}
Autonomous 6G network management requires agents that can execute tools, observe the resulting state changes, and adapt their decisions accordingly. Existing benchmarks based on static questions or scripted episode replay, however, do not support such closed-loop interaction, limiting agents to passive evaluation without the ability to learn from environmental feedback. This paper presents \textbf{6GAgentGym} to provide closed-loop capability. The framework provides an interactive environment with 42 typed tools whose effect classification distinguishes read-only observation from state-mutating configuration, backed by a learned Experiment Model calibrated on NS-3 simulation data. \textbf{6G-Forge} bootstraps closed-loop training trajectories from NS-3 seeds via iterative Self-Instruct generation with execution verification against the Experiment Model. Supervised fine-tuning on the resulting corpus followed by reinforcement learning with online closed-loop interaction enables an 8B open-source model to achieve comparable overall success rate to GPT-5 on the accompanying \textbf{6GAgentBench}, with stronger performance on long-horizon tasks. Together, these components provide a viable path toward autonomous, closed-loop network management.
\end{abstract}


\section{Introduction}
\label{sec:introduction}

The convergence of 6G with massive IoT introduces substantial complexity in network management, spanning edge intelligence, dynamic slicing, and space-air-ground integration~\cite{nguyen2021_6g_iot}. Configuring and optimizing a modern network that simultaneously manages intent translation, dynamic network slicing, trust policies, and multi-agent coordination typically requires an experienced engineer hours to days of manual effort per deployment. Industry analyses indicate that 45\% of network outages stem from configuration and change management failures. As the number of radio-access technologies, band combinations, and network slices grows, manual optimization becomes increasingly difficult. While deep learning has improved network traffic prediction~\cite{aouedi2025_traffic} and multi-level deep RL has shown promise for automated slicing in O-RAN-based 6G networks~\cite{ghafouri2024_oran_slicing}, these approaches address isolated subtasks and do not yet provide the end-to-end, semantically grounded management that operators need.

\begin{figure}[!t]
    \centering
    \includegraphics[width=1\columnwidth]{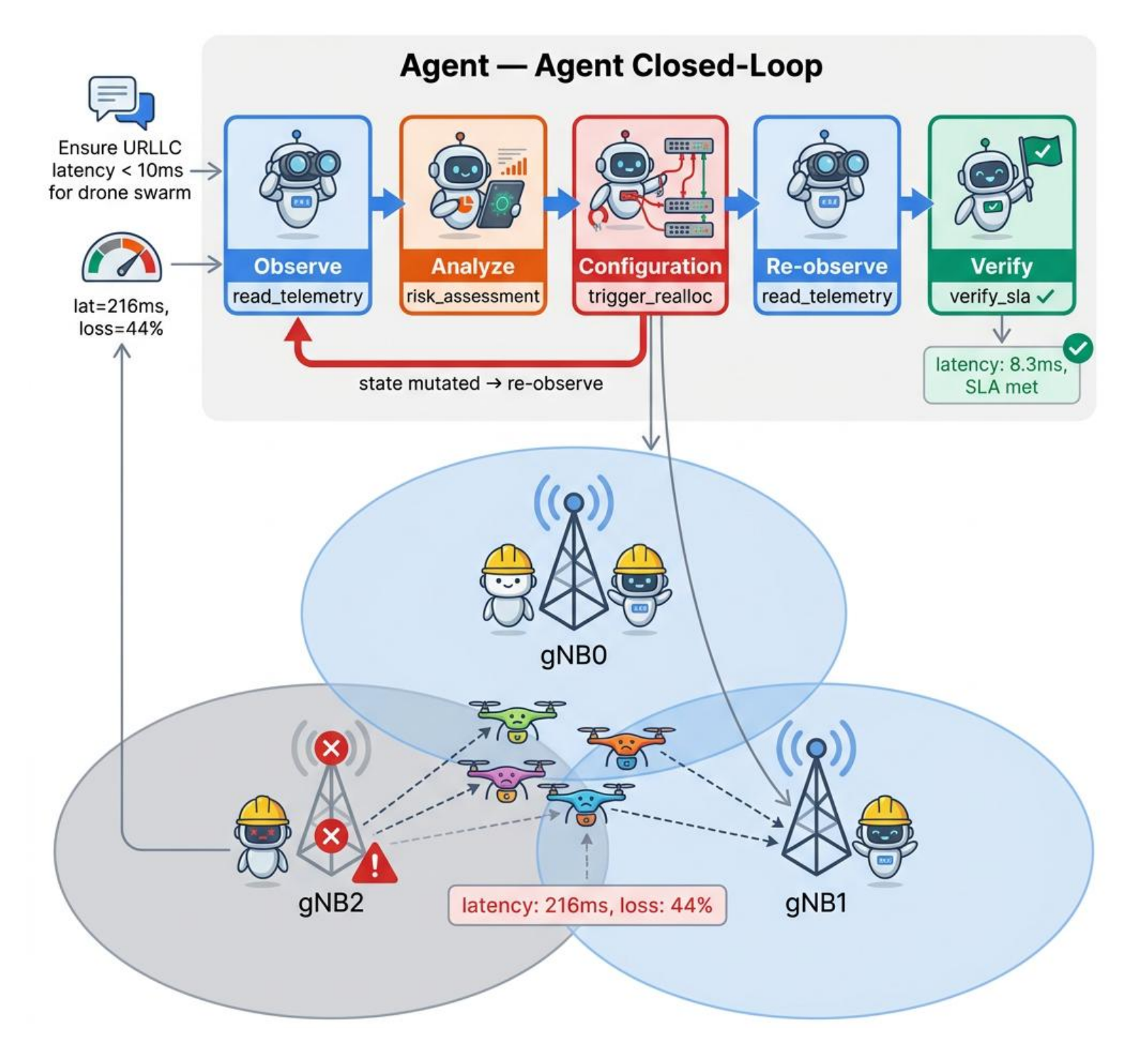}
    \caption{Agent interaction model. The agent selects tools in a closed loop until the operator intent is verified. Configuration tools mutate network state, requiring re-observation.}
    \label{fig:agent_loop}
\end{figure}

Large language models (LLMs) offer a promising path toward autonomous network management. Their capabilities in multi-step planning and tool orchestration~\cite{schick2023toolformer,wu2024netllm} allow agents to interpret operator intent in natural language, select appropriate network functions, and adapt decisions based on execution feedback~\cite{alpha3bench,maatouk2023llmtelecom,zhou2024llmtelecom}. Recent work has explored LLM-driven digital twins for closed-loop network optimization~\cite{guo2026_llm_dt} and LLM-based intent translation for intent-based networking~\cite{su2025_intent_ibn}, demonstrating the breadth of LLM applications in network management. Fig.~\ref{fig:agent_loop} illustrates this vision: the agent receives an operator intent, repeatedly selects and executes typed tools against the network, observes state changes, and continues until the intent is verified. However, network agents differ fundamentally from general-purpose LLM agents: they act on physical infrastructure under hard latency, safety, and regulatory constraints, demanding that such constraints be embedded in the agent's action space and decision logic rather than treated as post-hoc filters. Realizing this closed-loop interaction at scale therefore requires answering a fundamental question: how can we build the interactive environment, synthesize closed-loop training data, and train agents so that compact models reliably manage 6G networks?

Several lines of work provide building blocks toward this goal. In the general agent community, WebArena~\cite{zhou2024webarena} and SWE-bench~\cite{jimenez2024swebench} demonstrate the value of interactive, execution-grounded evaluation, but target web navigation and software engineering rather than network operations. In the 6G domain, 6G-Bench~\cite{6gbench} evaluates models through static multiple-choice questions without tool execution. $\alpha^3$-Bench~\cite{alpha3bench} and $\alpha^3$-SecBench~\cite{alpha3secbench} construct large-scale conversational episodes with tool calls under dynamic 6G conditions, but episodes are scripted replays rather than interactive environments. The scaling study in~\cite{tiny6g} reveals favorable edge-deployment trade-offs for mid-scale models, yet remains limited to static reasoning. On the tool-use training side, general-purpose data synthesis pipelines such as APIGen~\cite{liu2024apigen} and DIVE~\cite{dive2026} produce verifiable function-calling datasets but treat tools as stateless transformations without environmental feedback. Traditional deep RL approaches for network slicing~\cite{li2024drl_slicing,li2019drl_slicing_seminal} do operate in closed-loop environments but at the level of low-dimensional state-action spaces, lacking the semantic reasoning that LLM agents provide. What remains underexplored is an integrated platform where agents interact with a network environment in closed loop---executing state-mutating tools, observing consequences, and learning from the resulting trajectories.

This paper presents \textbf{6GAgentGym}, a framework that unifies interactive environment, data synthesis, and agent training for 6G network management. The contributions are:

\begin{itemize}
    \item \textbf{Interactive closed-loop environment.} We construct an execution platform with 42 typed tools classified by effect (Observation, Reasoning, Configuration), backed by a learned Experiment Model calibrated on NS-3 simulation data that predicts six-dimensional network state transitions.
    \item \textbf{Scalable data synthesis.} 6G-Forge bootstraps training trajectories from NS-3 seed data through iterative Self-Instruct generation constrained by typed tool signatures and effect labels, with execution verified against the Experiment Model and diversity maintained through ROUGE-L deduplication.
    \item \textbf{Agent training and evaluation.} Agentic SFT followed by RL with online closed-loop interaction enables an 8B model to achieve comparable overall performance to GPT-5 on 6GAgentBench (L1--L3), with an advantage on long-horizon L3 tasks.
\end{itemize}

\section{Related Work}
\label{sec:related_work}

\subsection{LLM Agents for 6G Networks}

Evaluating whether LLM can handle 6G network management tasks has attracted growing attention. 6G-Bench~\cite{6gbench} establishes a standardization-aligned taxonomy of 30 decision-making tasks and evaluates 22 foundation models on 3,722 multiple-choice questions. $\alpha^3$-Bench~\cite{alpha3bench} moves beyond static QA by formulating UAV missions as multi-turn conversational control loops with MCP and A2A tool calls, constructing 113k episodes and evaluating 17 models. $\alpha^3$-SecBench~\cite{alpha3secbench} augments this corpus with 20,000 adversarial scenarios. The scaling study in~\cite{tiny6g} shows that 1.5--3B models achieve the best edge-deployment trade-off on 6G-Bench. Beyond benchmarking, Guo et al.~\cite{guo2026_llm_dt} survey LLM-driven digital twins for network optimization, showing that LLMs can leverage real-time twin data to generate optimization strategies through closed-loop feedback. Su et al.~\cite{su2025_intent_ibn} demonstrate that LLMs with submodular in-context learning can translate high-level operator intents into executable network policies, achieving notable accuracy gains without parameter updates. Despite this progression from static evaluation to LLM-driven optimization, existing work either relies on pre-recorded episodes or addresses individual subtasks; to our knowledge, no existing agent executes a tool that mutates network state and adapts to the consequences within a unified interactive environment.

\begin{figure*}[!t]
    \centering
    \includegraphics[width=1\textwidth]{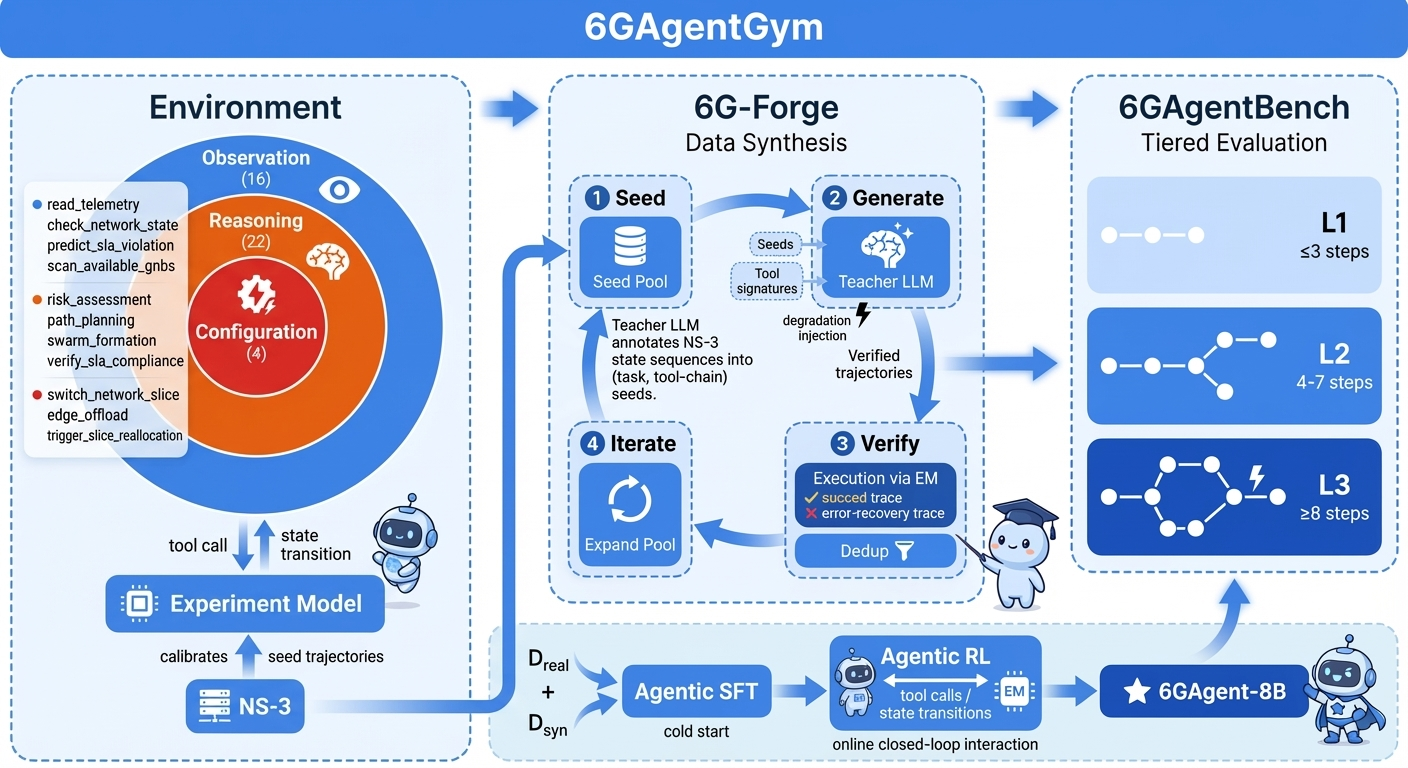}
    \caption{Overview of the 6GAgentGym framework. \textbf{Left}: interactive environment with 42 typed tools and Experiment Model. \textbf{Center}: 6G-Forge bootstraps closed-loop trajectories via iterative Self-Instruct generation with execution verification. \textbf{Right}: 6GAgentBench tiered evaluation (L1--L3). Below: Agentic SFT + RL training pipeline.}
    \label{fig:overview}
\end{figure*}

\subsection{Tool-Use Benchmarks and Data Synthesis}

Training LLM agents to use tools reliably requires both evaluation environments and scalable training data. Early benchmarks such as ToolLLM~\cite{qin2023toolllm}, Gorilla~\cite{patil2024gorilla}, and API-Bank~\cite{li2023apibank} evaluate API call generation over broad catalogs, but operate on stateless REST calls where tool outputs are independent of prior actions. Interactive benchmarks address this limitation: WebArena~\cite{zhou2024webarena} provides a realistic web environment for long-horizon tasks, SWE-bench~\cite{jimenez2024swebench} evaluates agents on real GitHub issues, and AgentBoard~\cite{ma2024agentboard} offers fine-grained progress metrics. These establish execution-grounded evaluation as a principle but do not target network operations.

On the data synthesis side, generating high-quality tool-use trajectories at scale remains an open problem. Self-Instruct~\cite{wang2023selfinstruct} bootstraps instruction data from a small seed set through iterative LLM generation. APIGen~\cite{liu2024apigen} extends this to function-calling datasets with three-stage verification, and DIVE~\cite{dive2026} demonstrates that diversity scaling improves agentic generalization. Domain-specific work on typed tool action spaces~\cite{shen2026sciagentgym} shows that structured training data can enable small models to outperform larger baselines. These pipelines, however, remain open-loop: tools transform data along acyclic chains without execution feedback altering environment state.

\subsection{Reinforcement Learning for Network Management}

Applying RL to network management has a long history, but the action spaces differ fundamentally from LLM tool use. Deep RL has been applied to network slicing resource allocation~\cite{li2019drl_slicing_seminal,li2024drl_slicing}; more recently, Ghafouri et al.~\cite{ghafouri2024_oran_slicing} propose a multi-level deep RL framework for O-RAN-based 6G cell-free networks that combines centralized multi-agent decision-making with decentralized execution. Deep learning has also been applied to network traffic prediction~\cite{aouedi2025_traffic}, providing complementary forecasting capabilities. These methods optimize over fixed, low-dimensional action spaces (e.g., bandwidth allocation per slice) and lack the ability to interpret operator intent or compose multi-step diagnostic workflows. LLM-based agents offer a complementary paradigm at the semantic level: the ReAct framework~\cite{yao2023react} interleaves reasoning traces with tool actions, while the World Knowledge Model~\cite{song2024wkm} shows that LLM agents can leverage both prior and dynamic state knowledge to mitigate blind trial-and-error. More recently, SWE-World~\cite{sweworld2025} replaces containerized execution environments with learned surrogate models trained on real interaction data, enabling Docker-free agent training and test-time scaling for software engineering tasks. In the networking domain, NS-3~\cite{riley2010ns} remains the standard discrete-event simulator for protocol-level research, with recent extensions integrating ray-tracing channels for 6G multi-RAT scenarios~\cite{pegurri2025toward}; our work adopts NS-3 as the physical-layer ground truth and distills its dynamics into a learned surrogate to enable scalable agent training.

\section{Methodology}
\label{sec:method}

Fig.~\ref{fig:overview} provides an overview of the framework. The following subsections detail each component.

\subsection{Agent Interaction Model}
\label{sec:interaction}

The network management agent operates as a closed-loop controller (Fig.~\ref{fig:agent_loop}). At each step, the agent receives the current network state $\mathbf{n}_t$ and the operator's intent $Q$ (expressed in natural language), selects a typed tool $v_t \in \mathcal{V}$ with arguments $\mathbf{x}_t$, and submits the call to the Experiment Model $M_{\theta}$. The environment returns a tool result $o_t$ and an updated state $\mathbf{n}_{t+1}$; the agent appends this observation to its history and decides the next action. The loop terminates when the agent invokes a verification tool confirming that the intent has been satisfied, or when a maximum step budget is reached.

This interaction model differs from static QA evaluation (where no tool is executed), scripted replay (where tool results are pre-recorded), and low-dimensional RL (where actions are continuous vectors without semantic structure).

\begin{figure*}[!t]
    \centering
    \includegraphics[width=1\textwidth]{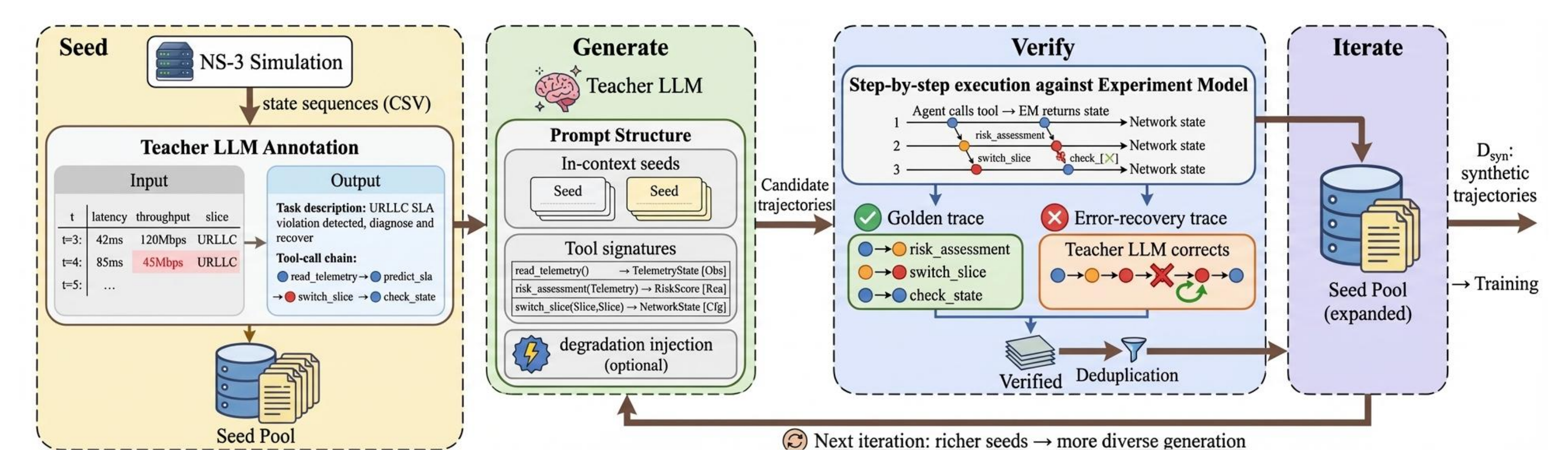}
    \caption{The 6G-Forge data synthesis pipeline. \textbf{Step 1}: NS-3 traces are annotated into seed trajectories. \textbf{Step 2}: A teacher LLM generates new trajectories from seed demonstrations. \textbf{Step 3}: Execution against $M_{\theta}$ produces golden and error-recovery traces. \textbf{Step 4}: Verified trajectories expand the seed pool; Steps 2--4 repeat for $K$ iterations.}
    \label{fig:pipeline}
\end{figure*}

\subsection{Interactive Environment}
\label{sec:environment}

\subsubsection{Typed Tool System}

We formalize the tool space as a finite set $\mathcal{V}$ of 42 typed functions over a domain type hierarchy $\mathcal{T}$ (see Appendix~\ref{app:tools} for the full catalog). Each tool is classified by its effect on the network state $\mathbf{n}_t$, and $\mathcal{V}$ is partitioned into three disjoint subsets accordingly:
\begin{equation}
    \mathcal{V} = \mathcal{V}_{\text{obs}} \;\sqcup\; \mathcal{V}_{\text{rea}} \;\sqcup\; \mathcal{V}_{\text{cfg}}
\end{equation}

$\mathcal{V}_{\text{obs}}$ (\textbf{Observation}) provides read-only access to the network state without modifying $\mathbf{n}_t$. $\mathcal{V}_{\text{rea}}$ (\textbf{Reasoning}) performs pure computation on typed inputs, neither reading nor modifying $\mathbf{n}_t$. $\mathcal{V}_{\text{cfg}}$ (\textbf{Configuration}) contains state-mutating operations where $\mathbf{n}_{t+1} \neq \mathbf{n}_t$ in general; these are the only tools that can invalidate prior observations, requiring re-observation to close the feedback loop. The tool space spans both network management and UAV control, forcing the agent to learn cross-domain trade-offs~\cite{alpha3bench}.

\subsubsection{NS-3 Simulation Backend}

The physical-layer foundation of 6GAgentGym is NS-3~\cite{riley2010ns}, the widely adopted discrete-event network simulator that provides protocol-level fidelity, modular architecture, and reproducible experimentation. We build on its NR module for 5G New Radio air-interface modeling and FlowMonitor for per-flow metrics extraction, enabling us to capture the six-dimensional state dynamics (slice type, latency, jitter, loss, throughput, edge load) that underpin both environment calibration and trajectory generation. NS-3 serves two roles in our framework: (1)~generating the seed traces from which the Experiment Model and 6G-Forge originate, and (2)~providing the ground-truth simulator for final policy validation (\S\ref{sec:ns3_validation}).

\subsubsection{Experiment Model}

Both trajectory synthesis and RL training require the agent to interact with an environment over many steps. Running full-fidelity NS-3 simulation for every such interaction is prohibitively expensive. We therefore introduce the Experiment Model $M_{\theta}$, a compact LLM fine-tuned on NS-3 traces to serve as a learned environment proxy. Rather than replicating the simulator exactly, $M_{\theta}$ learns state transitions that are sufficiently consistent and causally grounded to support downstream training. It operates in a six-dimensional network state space
\begin{equation}
    \mathbf{n}_t = (s_t,\, \ell_t,\, j_t,\, \rho_t,\, \tau_t,\, e_t) \;\in\; \mathcal{N}
\end{equation}
where $s_t$ is the active slice type, $\ell_t$ latency, $j_t$ jitter, $\rho_t$ loss rate, $\tau_t$ throughput, and $e_t$ edge-compute load.

\textbf{Inference.} At each step, $M_{\theta}$ receives the current state $\mathbf{n}_t$, the agent's tool call $(v_t, \mathbf{x}_t)$, an optional degradation event $\delta_t$, and the interaction history $\{(\mathbf{n}_i, v_i, \mathbf{x}_i)\}_{i<t}$. It produces a chain-of-thought reasoning trace explaining how the tool call affects the network, followed by the predicted next state and tool result:
\begin{equation}
\label{eq:exp_model}
    (\hat{\mathbf{n}}_{t+1},\, \hat{o}_t) = M_{\theta}(\mathbf{n}_t,\, v_t,\, \mathbf{x}_t,\, \delta_t,\, H_t)
\end{equation}
For Observation and Reasoning tools without degradation, $M_{\theta}$ preserves state; for Configuration tools, it predicts the corresponding transition dynamics.

\textbf{Training.} $M_{\theta}$ is trained via supervised fine-tuning on transition tuples extracted from NS-3 traces, with a joint objective over reasoning generation and next-state prediction:
\begin{equation}
\label{eq:exp_train}
    \mathcal{L} = -\mathbb{E}\big[\log P_{\theta}(\mathbf{n}_{t+1} \mid \mathbf{n}_t, v_t, \mathbf{x}_t, H_t) + \log P_{\theta}(o_t \mid \mathbf{n}_t, v_t, \mathbf{x}_t, H_t)\big]
\end{equation}
This objective ensures that $M_{\theta}$ learns both to predict consistent state transitions and to generate faithful tool return values.

\subsection{6G-Forge: Data Synthesis}
\label{sec:forge}


Training an agent to manage 6G networks requires large-scale trajectory data, yet three barriers obstruct its collection: real NS-3 rollouts are computationally expensive, hand-crafting multi-step tool sequences demands deep domain expertise, and static datasets lack the diversity needed for robust generalization. Inspired by Self-Instruct~\cite{wang2023selfinstruct}, 6G-Forge addresses all three by bootstrapping a trajectory corpus from a small set of NS-3 seeds through iterative LLM-driven generation with execution verification against $M_{\theta}$ (Fig.~\ref{fig:pipeline}). The pipeline has four steps.

\subsubsection{Step 1: Seed Annotation}

NS-3 simulations produce network state time series. These raw traces do not contain tool-call sequences. A teacher LLM converts each trace into a seed trajectory by identifying decision points (service-level agreement (SLA) violations, handover events, degradation onsets) and generating the corresponding task description $Q_i$ and tool-call sequence $\boldsymbol{\xi}_i = [(v_j, \mathbf{x}_j, o_j, \mathbf{n}_j)]_{j=1}^{L_i}$. The resulting seed pool $\mathcal{P}^{(0)} = \mathcal{D}_{\text{real}}$ is grounded in physical-layer dynamics and serves as the quality anchor for subsequent generation.

\subsubsection{Step 2: Self-Instruct Expansion}

At each iteration $k$, the teacher LLM receives a random sample of seed trajectories from $\mathcal{P}^{(k)}$ as in-context demonstrations, together with the typed tool signatures of $\mathcal{V}$. The teacher generates a new task description $Q'$ and a corresponding tool-call trajectory $\boldsymbol{\xi}'$. With probability $p_{\text{deg}}$, a degradation scenario is additionally specified, requiring the trajectory to include a recovery sub-sequence.

\subsubsection{Step 3: Execution Verification}

Each candidate trajectory is executed step-by-step against $M_{\theta}$ via Eq.~\eqref{eq:exp_model}. The teacher LLM's imagined state values are replaced by $M_{\theta}$'s predictions. Successful executions yield golden traces. When $M_{\theta}$ returns an error, the teacher is re-prompted to correct the failed step, producing an error-recovery augmented trajectory:
\begin{equation}
    \boldsymbol{\xi}^{*}_{\text{aug}} = [\ldots,\, (v_i, \mathbf{x}_i, e_i, \mathbf{n}_i),\, (v_i, \mathbf{x}_i', o_i, \mathbf{n}_i'),\, \ldots]
\end{equation}
Both golden and error-recovery trajectories are retained as training data.

\subsubsection{Step 4: Deduplication and Iterative Growth}

Each verified trajectory is compared against $\mathcal{P}^{(k)}$ using ROUGE-L; near-duplicates are discarded. Remaining trajectories are added to the pool: $\mathcal{P}^{(k+1)} = \mathcal{P}^{(k)} \cup \mathcal{B}^{(k)}$. Steps 2--4 repeat for $K$ iterations, producing the synthetic dataset $\mathcal{D}_{\text{syn}}$.

\subsection{6GAgentBench: Tiered Evaluation}
\label{sec:benchmark}

The evaluation suite is organized into three difficulty tiers aligned with 6G-Bench~\cite{6gbench}. \textbf{L1} ($\leq 3$ steps) covers elementary sense-decide-act chains. \textbf{L2} (4--7 steps) covers network-adaptive workflows such as degradation detection and slice reallocation. \textbf{L3} ($\geq 8$ steps) covers long-horizon multi-agent workflows under network degradation. Candidate tasks where any single model achieves $>$80\% zero-shot success are excluded.

\subsection{Two-Stage Agentic Training}
\label{sec:training}

\subsubsection{Stage 1: Agentic SFT}

The SFT dataset combines NS-3 real trajectories $\mathcal{D}_{\text{real}}$ and synthetic trajectories $\mathcal{D}_{\text{syn}}$ from 6G-Forge. The student model is fine-tuned on their union $\mathcal{D}_{\text{sft}} = \mathcal{D}_{\text{syn}} \cup \mathcal{D}_{\text{real}}$.

\subsubsection{Stage 2: Agentic RL}
\label{sec:rl}

Starting from the SFT checkpoint, the agent is further optimized through reinforcement learning with online tool interaction against $M_{\theta}$. Frontier tasks are selected by retaining only those where the SFT policy's empirical success rate falls within a learnable range. The composite reward is:
\begin{equation}
    R = \lambda \cdot R_{\text{format}} + R_{\text{correct}}
\end{equation}
where $R_{\text{format}}$ penalizes malformed tool calls and $R_{\text{correct}}$ measures task correctness. The policy is optimized with DAPO~\cite{dapo}.

\section{Experiments}
\label{sec:experiments}

\subsection{Experimental Setup}

\subsubsection{Models}
We evaluate eight frontier models spanning proprietary and open-source families: GPT-5, Claude-Sonnet-4, Gemini-2.5-Pro, Qwen3-VL-72B-Instruct, Qwen3-VL-8B-Instruct, Qwen3-VL-4B-Instruct, DeepSeek-V3, and Llama-4-Scout, along with three non-LLM baselines and our fine-tuned 6GAgent-8B and 6GAgent-4B (initialized from Qwen3-VL-8B-Instruct and Qwen3-VL-4B-Instruct, respectively). All LLM models are evaluated under ReAct-style~\cite{yao2023react} interaction loops with deterministic decoding (temperature 0).

\subsubsection{Training Data}
The SFT dataset combines two sources: (1)~3{,}000 real trajectories from NS-3 simulations (L1--L3), which also serve as the seed pool $\mathcal{P}^{(0)}$ for 6G-Forge; and (2)~50{,}000 synthetic trajectories from $K=15$ offline Self-Instruct iterations (30{,}000 golden traces + 20{,}000 error-recovery augmented traces), each verified through type checking, execution against $M_{\theta}$, and diversity filtering. Together these 53{,}000 trajectories span all five evaluation domains (network slicing, edge offloading, UAV control, degradation recovery, and multi-agent coordination) with controlled difficulty distribution: 30\% L1, 45\% L2, 25\% L3. For RL, approximately 8{,}000 frontier tasks are selected from this pool based on empirical success rate.

\subsubsection{Holdout Protocol}
To prevent data leakage between training and evaluation, we enforce a strict three-way holdout design across the tightly coupled environment--synthesis--training--evaluation pipeline. (1)~\textbf{Task-level holdout}: the 6GAgentBench evaluation tasks are drawn from a separate set of NS-3 scenario configurations (distinct topology seeds, traffic mixes, and failure injection patterns) that are never used during 6G-Forge synthesis or SFT/RL training. Specifically, the three NS-3 scenario scripts used for seed trajectory generation employ topology seeds 1--50, while the benchmark uses seeds 51--80. (2)~\textbf{Trajectory-level deduplication}: all evaluation task descriptions are compared against the 53k training corpus using ROUGE-L; any evaluation task with ROUGE-L~$\geq 0.7$ to any training instance is replaced with a manually curated alternative. This filtering removes 127 candidate tasks (8.4\% of the initial evaluation pool). (3)~\textbf{Experiment Model isolation}: the Experiment Model $M_{\theta}$ is frozen before RL training begins and is not updated based on evaluation-time interactions, ensuring that the surrogate dynamics cannot be tuned to favor benchmark-specific patterns. A TF-IDF separability probe achieves only 53.2\% accuracy (near chance) at distinguishing training vs.\ evaluation tasks.

\subsubsection{Metrics}
We adopt two primary metrics: \textbf{Success Rate (SR)}, the fraction of tasks where the agent produces a correct final answer through valid tool-use chains; and \textbf{Success weighted by Path Length (SPL)}, which penalizes unnecessarily long trajectories.

\subsection{Main Results}

Table~\ref{tab:main_results} presents overall performance across the three difficulty tiers.

\begin{table}[!t]
\centering
\footnotesize
\renewcommand\arraystretch{1.15}
\setlength{\tabcolsep}{1.2mm}
\caption{\textbf{Overall Success Rate (\%) on 6GAgentBench.} Best in \textbf{bold}, second best \underline{underlined}. $^\dagger$: our fine-tuned models.}
\label{tab:main_results}
\resizebox{\columnwidth}{!}{%
\begin{tabular}{lcccccc}
\toprule[1.2pt]
\multirow{2}{*}{\textbf{Model}} & \multirow{2}{*}{\textbf{Params}} & \multicolumn{3}{c}{\textbf{Success Rate $\uparrow$ (\%)}} & \multirow{2}{*}{\textbf{SR $\uparrow$}} & \multirow{2}{*}{\textbf{SPL $\uparrow$}} \\
\cmidrule(lr){3-5}
 & & \textbf{L1} & \textbf{L2} & \textbf{L3} & & \\
\midrule[0.8pt]
\midrule[0.8pt]
\rowcolor{gray!8}
\multicolumn{7}{l}{\textit{Proprietary Models}} \\
GPT-5              & --       & \textbf{72.4} & \textbf{51.3} & \underline{33.8} & \textbf{50.2} & \underline{42.7} \\
Claude-Sonnet-4    & --       & \underline{68.9} & 46.7 & 28.5 & 45.8 & 38.9 \\
Gemini-2.5-Pro     & --       & 65.3 & 44.2 & 30.1 & 44.1 & 37.2 \\
DeepSeek-V3        & 685B$^*$ & 63.8 & 42.5 & 27.6 & 42.3 & 35.8 \\
\midrule[0.5pt]
\rowcolor{gray!8}
\multicolumn{7}{l}{\textit{Non-LLM Baselines}} \\
Threshold-Rule     & --       & 51.3 & 18.6 & 4.2  & 22.1 & 19.8 \\
MAPE-K Heuristic   & --       & 54.8 & 25.7 & 8.9  & 27.5 & 23.4 \\
DRL-Slicing~\cite{li2024drl_slicing} & -- & 38.2 & 27.3 & 11.5 & 24.1 & 20.6 \\
\midrule[0.5pt]
\rowcolor{gray!8}
\multicolumn{7}{l}{\textit{Open-Source Models}} \\
Qwen3-VL-72B       & 72B      & 58.2 & 36.4 & 22.3 & 36.8 & 30.5 \\
Llama-4-Scout      & 109B$^*$ & 55.6 & 33.8 & 19.7 & 34.2 & 28.1 \\
Qwen3-VL-8B        & 8B       & 42.1 & 24.6 & 12.8 & 24.9 & 19.7 \\
Qwen3-VL-4B        & 4B       & 36.5 & 19.3 & 9.4  & 20.3 & 15.6 \\
\midrule[0.5pt]
\rowcolor{blue!5}
\multicolumn{7}{l}{\textit{SFT on 6G-Forge}} \\
\rowcolor{blue!5}
6GAgent-8B$^\dagger$       & 8B & 64.7 & 43.8 & 34.2 & 45.3 & 38.4 \\
\rowcolor{blue!5}
6GAgent-4B$^\dagger$       & 4B & 56.3 & 36.1 & 25.7 & 37.5 & 31.2 \\
\midrule[0.5pt]
\rowcolor{blue!10}
\multicolumn{7}{l}{\textit{SFT + RL on 6G-Forge}} \\
\rowcolor{blue!10}
6GAgent-8B$^\dagger$ +RL   & 8B & 68.2 & \underline{48.5} & \textbf{39.1} & \underline{50.1} & \textbf{42.8} \\
\rowcolor{blue!10}
6GAgent-4B$^\dagger$ +RL   & 4B & 60.8 & 40.7 & 30.4 & 42.0 & 35.6 \\
\bottomrule[1.2pt]
\multicolumn{7}{l}{\scriptsize $^*$Mixture-of-Experts total parameters.}
\end{tabular}%
}
\end{table}

\textbf{Non-LLM baselines.} We include three non-LLM baselines to contextualize the difficulty of 6GAgentBench beyond LLM-based approaches. \textit{Threshold-Rule} implements a fixed-threshold remediation policy: if latency exceeds the SLA bound, switch to URLLC; if throughput drops below 10\,Mbps, trigger edge offload. \textit{MAPE-K Heuristic} extends this with a Monitor--Analyze--Plan--Execute--Knowledge loop that maintains a lookup table of 50 hand-crafted condition$\to$action rules covering common degradation patterns. \textit{DRL-Slicing}~\cite{li2024drl_slicing} is a prediction-aided deep RL agent originally designed for online power allocation and user admission in RAN slicing, adapted to our environment by mapping its discrete admission and continuous power actions to the nearest Configuration tool calls.
All three baselines achieve reasonable L1 performance (38--55\%) but degrade sharply on L2/L3, indicating that the benchmark requires capabilities beyond what these fixed-policy and low-dimensional RL approaches can provide.

\textbf{Key findings:} (1)~All models exhibit sharp performance degradation from L1 to L3, indicating that multi-step tool-use remains a fundamental bottleneck. GPT-5 drops from 72.4\% to 33.8\%. (2)~6GAgent-8B (SFT+RL) achieves 50.1\% overall SR, comparable to GPT-5 (50.2\%) despite being an 8B open-source model, while substantially outperforming the larger Qwen3-VL-72B (36.8\%). (3)~On L3 tasks, 6GAgent-8B (SFT+RL) reaches 39.1\% versus GPT-5's 33.8\%, suggesting that closed-loop RL training is particularly beneficial for long-horizon tasks requiring degradation recovery. (4)~RL contributes +4.8\% overall on top of SFT, with larger gains on L3 (+4.9\%) than L1 (+3.5\%).

\subsection{Ablation Studies}

\subsubsection{Training Data Composition}

We distinguish two trajectory types: \emph{open-loop} trajectories, where tool outputs are pre-recorded and do not reflect state changes from prior actions; and \emph{closed-loop} trajectories, where each tool call executes against the Experiment Model and subsequent observations reflect the mutated state. Table~\ref{tab:ablation} incrementally adds training data components to isolate their contributions.

\begin{table}[!t]
\centering
\footnotesize
\renewcommand\arraystretch{1.2}
\setlength{\tabcolsep}{2.0mm}
\caption{\textbf{Ablation on Training Data Composition} (8B model). Each row adds one component; $\Delta$ shows incremental gain.}
\label{tab:ablation}
\begin{tabular}{lcccc>{\columncolor{gray!8}}c}
\toprule[1.2pt]
\textbf{Training Data} & \textbf{L1} & \textbf{L2} & \textbf{L3} & \textbf{SR $\uparrow$} & \textbf{$\Delta$} \\
\midrule[0.8pt]
\midrule[0.8pt]
Baseline (no fine-tuning)       & 42.1 & 24.6 & 12.8 & 24.9 & -- \\
Open-loop 6G trajectories       & 56.2 & 35.4 & 24.1 & 37.0 & \small+12.1 \\
Closed-loop synthetic only      & 61.3 & 40.2 & 29.6 & 41.8 & \small+4.8 \\
\quad + error recovery          & 62.8 & 41.5 & 31.4 & 43.1 & \small+1.3 \\
\rowcolor{blue!5}
\quad + NS-3 real data (SFT)    & 64.7 & 43.8 & 34.2 & 45.3 & \small+2.2 \\
\rowcolor{blue!10}
\quad + Agentic RL              & \textbf{68.2} & \textbf{48.5} & \textbf{39.1} & \textbf{50.1} & \small+4.8 \\
\bottomrule[1.2pt]
\end{tabular}
\end{table}

Each row adds one component; the incremental gains isolate individual contributions. Closed-loop trajectories add +4.8\% over open-loop, showing that capturing state mutations from Configuration tools matters. NS-3 real data adds +2.2\% overall (+2.8\% on L3), recovering physical-layer reasoning that synthetic generation abstracts away. Agentic RL adds +4.8\% on top of SFT, with disproportionate L3 gains (+4.9\% vs.\ +3.5\% on L1).

\subsection{Result Analysis}

\begin{figure*}[!t]
    \centering
    \includegraphics[width=0.9\textwidth]{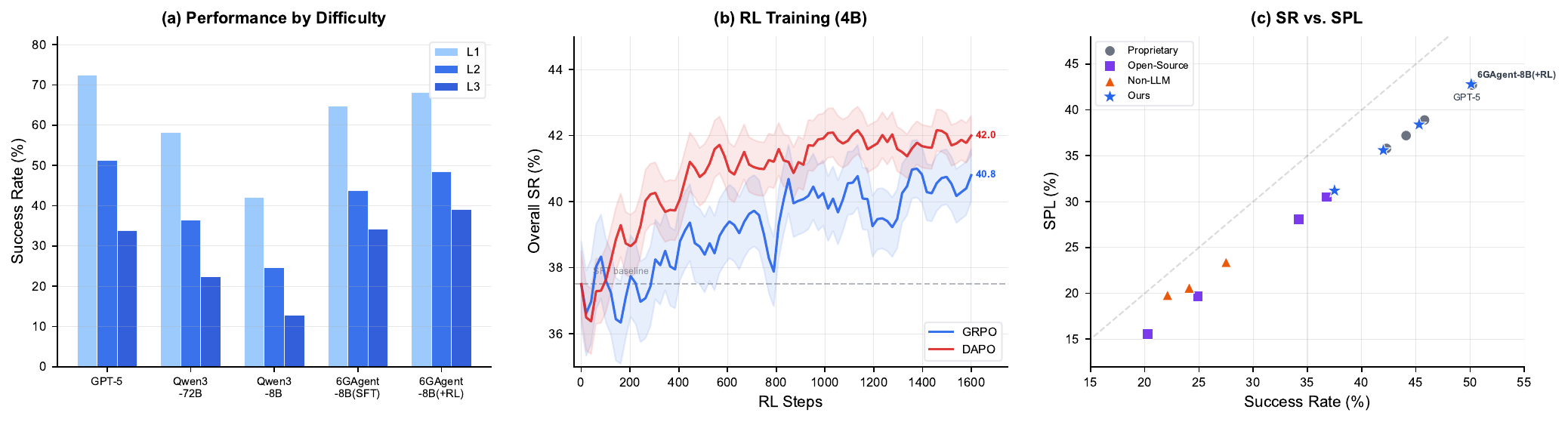}
    \caption{Visual analysis of 6GAgentBench results. \textbf{(a)}~Performance by difficulty tier. \textbf{(b)}~GRPO vs.\ DAPO RL training on the 4B model. \textbf{(c)}~SR vs.\ SPL.}
    \label{fig:scaling}
\end{figure*}

Fig.~\ref{fig:scaling} visualizes the main results from three perspectives.

\textbf{Difficulty-tier breakdown} (Fig.~\ref{fig:scaling}a). GPT-5 leads on L1 (72.4\% vs.\ 68.2\%), but 6GAgent-8B (+RL) surpasses it on L3 (39.1\% vs.\ 33.8\%), suggesting that closed-loop RL is particularly beneficial for long-horizon tasks. Base open-source models (Qwen3-8B) struggle across all tiers, highlighting the importance of domain-specific training.

\textbf{RL algorithm comparison} (Fig.~\ref{fig:scaling}b). We compare GRPO and DAPO on the 4B model over 1.6k RL steps. Both algorithms show rapid initial gains and gradual saturation, but DAPO converges faster and reaches a higher final SR (42.0\% vs.\ 40.8\%), benefiting from its dynamic sampling strategy that prioritizes under-explored task types. 
\textbf{Efficiency analysis} (Fig.~\ref{fig:scaling}c). 6GAgent-8B (+RL) achieves the highest SPL (42.8\%), slightly exceeding GPT-5 (42.7\%), indicating that RL improves both task completion and path efficiency.

\subsection{NS-3 Grounded Validation}
\label{sec:ns3_validation}

A central concern with surrogate-based evaluation is whether agent policies optimized against the Experiment Model $M_{\theta}$ transfer to the full-fidelity NS-3 simulator. To quantify this sim-to-sim gap, we replay the tool-call trajectories produced by each model on a held-out subset of 120 tasks (40 per tier) directly in NS-3, where each tool call triggers the corresponding NS-3 API and the resulting network state is measured from the simulator rather than predicted by $M_{\theta}$.

\begin{table}[!t]
\centering
\footnotesize
\renewcommand\arraystretch{1.2}
\setlength{\tabcolsep}{2.2mm}
\caption{\textbf{NS-3 Grounded Validation.} Success Rate (\%) under Experiment Model vs.\ full-fidelity NS-3 replay on 120 held-out tasks.}
\label{tab:ns3_validation}
\begin{tabular}{lcc>{\columncolor{gray!8}}c>{\columncolor{gray!8}}c}
\toprule[1.2pt]
\textbf{Model} & \textbf{$M_{\theta}$ $\uparrow$} & \textbf{NS-3 $\uparrow$} & \textbf{$\Delta$ $\downarrow$} & \textbf{$r$ $\uparrow$} \\
\midrule[0.8pt]
\midrule[0.8pt]
GPT-5              & 50.0 & 47.5 & $-$2.5 & 0.94 \\
Claude-Sonnet-4    & 45.8 & 43.3 & $-$2.5 & 0.93 \\
Qwen3-VL-72B       & 37.5 & 34.2 & $-$3.3 & 0.91 \\
Qwen3-VL-8B        & 25.0 & 22.5 & $-$2.5 & 0.92 \\
\midrule[0.5pt]
\rowcolor{blue!5}
6GAgent-8B (SFT)    & 45.0 & 42.5 & $-$2.5 & 0.93 \\
\rowcolor{blue!10}
6GAgent-8B (SFT+RL) & 50.8 & 47.5 & $-$3.3 & 0.92 \\
\midrule[0.5pt]
Threshold-Rule      & 21.7 & 21.7 & 0.0   & 1.00 \\
MAPE-K Heuristic    & 28.3 & 27.5 & $-$0.8 & 0.98 \\
\bottomrule[1.2pt]
\end{tabular}
\end{table}

Table~\ref{tab:ns3_validation} reports the results. Three findings emerge. (1)~\textbf{The surrogate gap is small and consistent}: across all models, the absolute SR drop from $M_{\theta}$ to NS-3 ranges from 0.0\% to 3.3\%, with a mean gap of 2.2\%. The Pearson correlation between $M_{\theta}$-evaluated and NS-3-evaluated SR across all models is $r = 0.99$ ($p < 0.001$), indicating that $M_{\theta}$ preserves the relative ranking of all methods. (2)~\textbf{RL gains transfer to NS-3}: the gap for 6GAgent-8B (SFT+RL) is comparable to the SFT-only variant (3.3\% vs.\ 2.5\%), and the RL advantage is preserved in NS-3 (47.5\% vs.\ 42.5\%, +5.0\%). (3)~\textbf{Gap concentrates on L3 degradation recovery}: per-tier analysis shows L1 gap below 1\%, L2 gap around 2\%, and L3 gap around 4\%, attributable to transient dynamics (queue draining, handover timing) that $M_{\theta}$ approximates but NS-3 simulates at protocol level. This validates $M_{\theta}$ as a faithful training proxy while identifying protocol-level transients as the primary fidelity bottleneck for future improvement.

\begin{figure*}[!]
    \centering
    \begin{subfigure}[t]{0.32\textwidth}
        \centering
        \includegraphics[width=\textwidth]{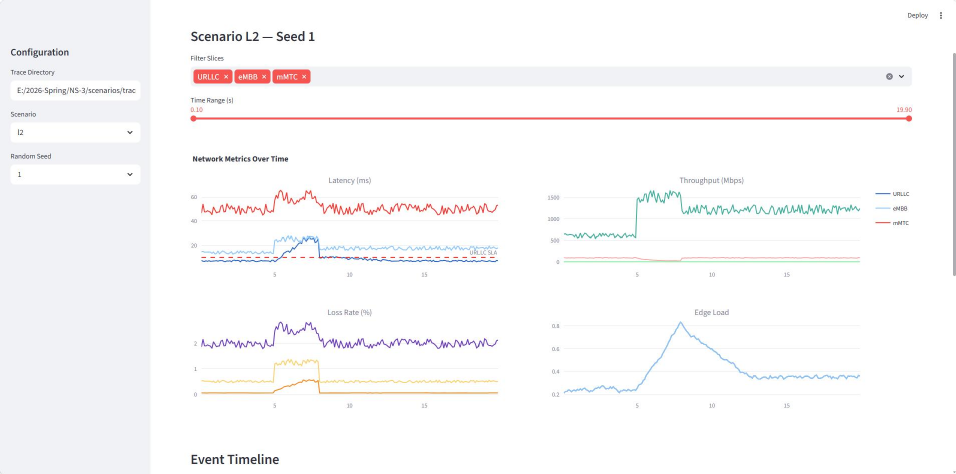}
        \caption{Trace Explorer.}
        \label{fig:dash_trace}
    \end{subfigure}
    \hfill
    \begin{subfigure}[t]{0.32\textwidth}
        \centering
        \includegraphics[width=\textwidth]{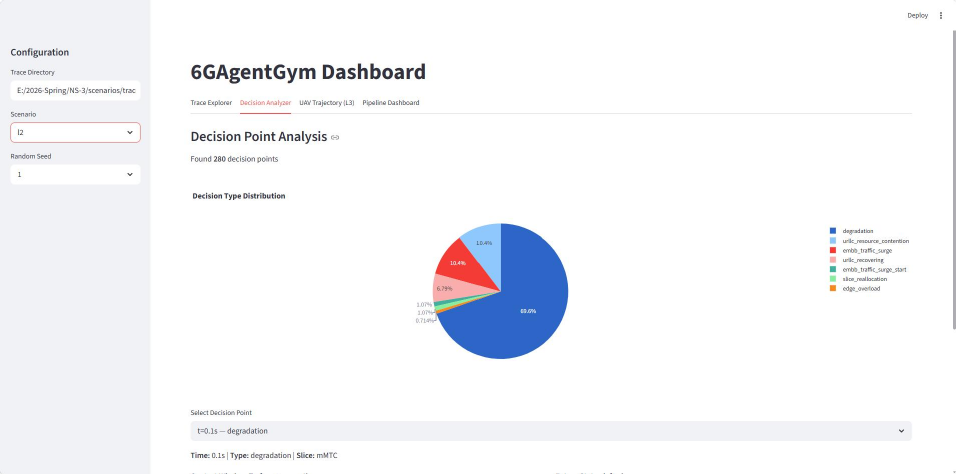}
        \caption{Decision Analyzer.}
        \label{fig:dash_decision}
    \end{subfigure}
    \hfill
    \begin{subfigure}[t]{0.32\textwidth}
        \centering
        \includegraphics[width=\textwidth]{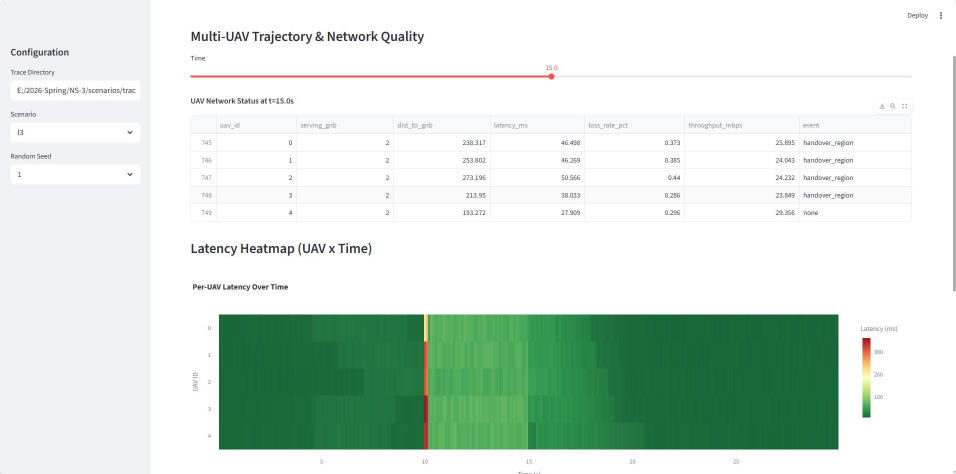}
        \caption{UAV Trajectory (L3).}
        \label{fig:dash_uav}
    \end{subfigure}
    \caption{6GAgentGym interactive visualization dashboard. (a)~Six-dimensional network metrics with SLA overlays. (b)~Decision point identification with type distribution. (c)~Per-UAV latency heatmap revealing handover-induced spikes.}
    \label{fig:dashboard}
\end{figure*}

\section{Conclusion}
\label{sec:conclusion}

This paper presents 6GAgentGym, a framework integrating an interactive environment with 42 typed tools and a learned Experiment Model, a Self-Instruct data synthesis pipeline (6G-Forge) seeded from NS-3 real data, and a tiered evaluation suite (6GAgentBench). Agentic SFT followed by RL with online closed-loop interaction enables an 8B model to achieve comparable overall performance to GPT-5, with an advantage on long-horizon tasks.

Several limitations remain before practical deployment. The Experiment Model approximates NS-3 dynamics but does not capture full protocol-level transients, particularly during handover and failure recovery. The current tool set of 42 covers network slicing and UAV control but excludes radio-level operations such as beamforming and power control. Extending RL to larger trajectory budgets and integrating multi-modal inputs (spectrum visualizations, topology maps) are directions for future work.

\bibliographystyle{IEEEtran}
\bibliography{references}

\appendices

\section{Full Tool Catalog}
\label{app:tools}

Table~\ref{tab:full_tools} lists all 42 tools with their typed signatures. Tools are grouped by effect category: Observation (read-only), Reasoning (pure computation), and Configuration (state-mutating).

\section{Interactive Visualization Dashboard}
\label{app:viz}

To support trajectory debugging and error analysis, we develop an interactive Streamlit-based dashboard that reads directly from NS-3 CSV traces. Fig.~\ref{fig:dashboard} illustrates its three primary views.

\begin{table}[!th]
\centering
\scriptsize
\setlength{\tabcolsep}{1.5mm}
\caption{Complete tool catalog of 6GAgentGym.}
\label{tab:full_tools}
\begin{tabular}{lll}
\toprule
\textbf{Tool Name} & \textbf{Input} & \textbf{Output} \\
\midrule
\multicolumn{3}{l}{\textbf{Observation} ($\mathcal{V}_{\text{obs}}$, 16 tools)} \\
\texttt{read\_telemetry} & -- & TelemetryState \\
\texttt{check\_network\_state} & -- & NetworkState \\
\texttt{get\_signal\_strength} & Position & SignalStrength \\
\texttt{scan\_available\_gnbs} & Position & GnbList \\
\texttt{get\_edge\_load} & -- & EdgeLoad \\
\texttt{get\_slice\_status} & SliceType & SliceStatus \\
\texttt{read\_uav\_position} & UavId & Position \\
\texttt{get\_battery\_level} & UavId & BatteryLevel \\
\texttt{predict\_sla\_violation} & NetworkState & SLAPrediction \\
\texttt{check\_handover\_status} & UavId & HandoverStatus \\
\texttt{get\_traffic\_pattern} & SliceType & TrafficPattern \\
\texttt{monitor\_interference} & Position & InterferenceLevel \\
\texttt{check\_link\_quality} & UavId, GnbId & LinkQuality \\
\texttt{select\_recovery\_strategy} & NetworkState, RiskScore & RecoveryStrategy \\
\texttt{get\_available\_slices} & Position & SliceList \\
\texttt{check\_migration\_feasib.} & UavId, GnbId & FeasibilityScore \\
\midrule
\multicolumn{3}{l}{\textbf{Reasoning} ($\mathcal{V}_{\text{rea}}$, 22 tools)} \\
\texttt{activate\_sensor} & SensorType & SensorHandle \\
\texttt{risk\_assessment} & TelemetryState & RiskScore \\
\texttt{evaluate\_intent\_feasib.} & Intent, NetworkState & FeasibilityScore \\
\texttt{check\_geofence} & Position, GeofenceSpec & GeofenceResult \\
\texttt{path\_planning} & Position$\times$2, NetworkState & Waypoints \\
\texttt{compute\_energy\_budget} & Waypoints, BatteryLevel & EnergyPlan \\
\texttt{select\_offload\_target} & EdgeLoad, TaskSpec & OffloadTarget \\
\texttt{negotiate\_priority} & UavId$\times$2, Intent & PriorityResult \\
\texttt{set\_waypoint} & UavId, Position & WaypointAck \\
\texttt{adjust\_altitude} & UavId, Altitude & AltitudeAck \\
\texttt{adjust\_speed} & UavId, Speed & SpeedAck \\
\texttt{collision\_avoidance} & UavId, Position, SwarmState & AvoidanceCmd \\
\texttt{swarm\_formation} & SwarmSpec, Waypoints & FormationCmd \\
\texttt{assign\_task} & UavId, TaskSpec & TaskAck \\
\texttt{send\_alert} & UavId, AlertType & AlertAck \\
\texttt{request\_handover} & UavId, GnbId & HandoverCmd \\
\texttt{log\_decision} & DecisionRecord & LogAck \\
\texttt{update\_mission\_plan} & MissionSpec, NetworkState & MissionPlan \\
\texttt{broadcast\_status} & UavId, StatusMsg & BroadcastAck \\
\texttt{heartbeat} & UavId & HeartbeatAck \\
\texttt{verify\_sla\_compliance} & NetworkState, SLASpec & ComplianceResult \\
\texttt{validate\_mission\_compl.} & MissionSpec, MissionLog & ValidationResult \\
\midrule
\multicolumn{3}{l}{\textbf{Configuration} ($\mathcal{V}_{\text{cfg}}$, 4 tools)} \\
\texttt{switch\_network\_slice} & SliceType$\times$2 & NetworkState \\
\texttt{graceful\_degradation} & DegradationSpec & NetworkState \\
\texttt{edge\_offload} & TaskSpec, OffloadTarget & OffloadResult \\
\texttt{trigger\_slice\_realloc.} & SliceType, ResourceSpec & NetworkState \\
\bottomrule
\end{tabular}
\end{table}

\end{document}